\documentclass[10pt]{article}
\usepackage[top=0.75in, bottom=0.75in, left=0.6in, right=0.6in]{geometry}
\usepackage{amsfonts,epsfig,graphicx,tabularx,amsmath,bm,relsize,steinmetz}
\usepackage{amsmath,epsfig,graphicx,setspace,url,float}
\usepackage{pstricks,pifont}
\usepackage{lscape,multicol,longtable}

\usepackage[inline]{enumitem}

\usepackage{cite}
\usepackage[english]{babel}
\usepackage[small,bf]{caption}
\usepackage{epstopdf}
\usepackage{multirow}
\usepackage[nameinlink]{cleveref}
\crefname{figure}{Fig.}{Figs.}
\crefname{table}{Table}{Tables}
\usepackage{wrapfig}
\usepackage{titlesec}
\titleformat*{\section}{\bfseries}
\titleformat*{\subsection}{\it}
\titleformat*{\subsubsection}{\it}

\title{Consensus of Multi-Agent Systems Using Back-Tracking and History Following Algorithms}
\author{Yanumula V. Karteek\thanks{Department of EEE, IIT Guwahati, Assam, India; email: {\textit {yanumula@alumni.iitg.ac.in, \{indranik,smajhi\}@iitg.ac.in}}} , \and Indrani Kar\footnotemark[1] , \and Somanath Majhi\footnotemark[1]}
\date{}
\begin {document}
\maketitle 

\section*{Abstract}
This paper proposes two algorithms, namely "back-tracking" and "history following", 
to reach consensus in case of communication loss for a network of distributed agents with 
switching topologies. To reach consensus in distributed control, considered communication 
topology forms a strongly connected graph. The graph is no more strongly connected whenever 
an agent loses communication.Whenever an agent loses communication, the topology 
is no more strongly connected. The proposed back-tracking algorithm 
makes sure that the agent backtracks its position unless the
communication is re-established, and path is changed to reach consensus. In history following, the agents 
use their memory and move towards previous consensus point until the communication 
is regained. Upon regaining communication, a new consensus point is calculated 
depending on the current positions of the agents and they change their trajectories 
accordingly. Simulation results, for a network of 
six agents, show that when the agents follow the previous history, the 
average consensus time is less than that of back-tracking. However, situation 
may arise in history following where a false notion of reaching consensus makes one of the agents 
stop at a point near to the actual consensus point. An obstacle
avoidance algorithm is integrated with the proposed algorithms to avoid collisions. Hardware implementation 
for a three robots system shows the effectiveness of the algorithms.

\section*{Key Words}
Consensus; networked multi-agent system; switching topologies based on position; back-tracking; history following


\section{Introduction}
The consensus problem was conceptualized after the exploration of self-ordered biologically motivated
particle motions by Vicsek
{\it et al.}  \cite{lt15} in 1995. A survey of the literature in this
area can be found in \cite{lt19}. When multiple
robots or agents are used to perform a combined task, they are
expected to cooperate with each other \cite{lt26,lt27,lt28}. To behave cooperatively, all
the agents must agree on a specific variable, be it position, velocity
or something else. When they concur on a suitable variable or
variables, the consensus is reached. Graph theory is popularly used to represent a network of 
agents with nodes and branches in a graph equivalent to agents and communication links 
among agents respectively \cite{lt17, lt16}. Saber and Murray \cite{lt17} considered 
undirected networks with distributed linear and nonlinear consensus protocols to solve an 
average consensus problem. They also found out an upper bound on
the tolerable fixed time-delay in the system. The same authors then
analyzed the convergence for three different cases in \cite{lt16}. 
Moreau \cite{lt09} investigated on the stability of consensus algorithms and derived sufficient conditions to
guarantee uniform exponential stability. Linear time varying systems
with communication time-delays are considered in this work and a
theoretical analysis was provided without any simulation results. Convergence of the topologies and performance are also 
addressed by them. Saber {\it et al.}
\cite{lt20}, provided a review of various first order consensus algorithms with and without time-delays. 
Sufficient condition for convergence in case of higher order systems is given by Ren \textit{et. al} \cite{lt03}. 
However, communication time-delays are not considered by the analysis.
A detailed study about time-delays in consensus protocols is presented in \cite{lt12}.  
Second order leaderless autonomous multi-agents with time-varying communication delays that are connected 
by a directed graph are considered in 
\cite{lt08} and the stability analysis is also provided. Directed graph with switching topologies and time-delays 
are considered in \cite{lt10} for first order with extension to higher order systems to prove the boundedness. 
A necessary and sufficient condition for consensus without time-delays in high-order systems with an
undirected fixed topology is presented in \cite{lt02}. Consensus of higher order multi-agents with fixed 
time-delays in discrete time is considered in \cite{lt01} using nearest-neighbour rule. 
A completely different approach was discussed in \cite{lt22} where a fuzzy sliding
mode controller was designed for consensus of multi-agent system without considering the time-delays.
Hardware implementation of multi-robot system for transportation is presented by Ting \textit{et. al} \cite{lt29}.

In this paper, a network of fifth order systems with switching
topologies and time-delays is considered in which some of them are not strongly
connected. Topology switching occurs in case of link loss and regain which is a very common phenomenon in any communication network. Presence of big 
obstacles like wall may cause loss of link and situation may arise when the link is never regained unless the position of the agent is changed. In such situations, positions of agents are closely related to switching of the communication topology. Two new algorithms called "back-tracking" and "history-following" are proposed to
efficiently handle such situations when the communication topology is not
strongly connected mainly because of the positions of the agents. 
According to the proposed back-tracking algorithm, an agent
backtracks its path if it loses communication with other agents and choose a new path to reach the consensus. 
However, in the history-following case, the agents use their memory to follow the previous 
consensus point once the communication is lost. A new consensus point is 
calculated after the communication is re-established and the network is strongly connected again. 
The group of $n$ agents and the underlying communication topology is represented by a directed graph with $n$ nodes 
and corresponding branches representing the communication topology. Branch weights of the directed graph are 
considered to be unity rather than non-uniform weights. The switching topologies based on position of agents are 
represented by corresponding directed graphs and adjacency matrices. Static obstacles are considered to be 
present in the paths of the agents which necessitates an obstacle avoidance algorithm to be incorporated with.

\section{System Model}\label{sysmod}
A group of $n$ robotic agents are considered with input and communication time-delays, network topology of the 
group of robotic agents is represented by a directed graph $G$\cite{lt16}. Then there exists a vertex set 
$ V = \{ 1,2,..,n\}\ $ representing nodes and an edge set $E = \{ (i,j):i,j \in V\}$ representing branches. 
The edge set comprises of ordered pairs $(i,j)$ if node $i$ is sending data to node $j$, which are represented 
by branches in the directed graph. The directed graph $G$ has an adjacency matrix $A = (a_{ji} )_{n \times n}\ $ 
with $a_{ii} = 0, \forall i \in V\ $, $a_{ji} = 1\ $ if $(i,j) \in E$ and $a_{ji} = 0\ $ otherwise. 
Communication time-delay among agents is represented by $g_{j}T$ and processing time-delay of each 
agent is represented by $g_{i}T$. The graph Laplacian matrix $L = (l_{ij})_{n \times n}$ of the directed graph $G$ 
have the elements $l_{ij} = \sum_{i \neq j} a_{ji}\ $ and $l_{ij} = -a_{ji}, \forall i \neq j$. For a strongly 
connected graph, $L$ has a zero eigenvalue and $[1,....,1]^T \in \Re^n\ $ is the corresponding eigenvector, rest of the eigenvalues have negative real parts.

\subsection{Simulation Model}

The $i^{th}$ agent dynamics in a coordinate system positioned at $\left( x_i, y_i\right) $ with an orientation of $\phi $radians is given by

\begin{equation}\label{eqn1}
\begin{array}{l}
 \dot x_i  = v_{i}\cos\left(\phi_{i}\right); \qquad \dot y_i  = v_{i}\sin\left(\phi_{i}\right) \\
 \dot \phi_{i} = \omega_{i}\; ; \qquad \dot v_i  = u_{i1}\; ;  \quad \dot \omega_{i} = u_{i2}
 \end{array}\
\end{equation}

where $v_i \in \Re^m $ and $\omega_i \in \Re^m $are linear velocity and angular velocity of the 
agent, $\dot v_i $ and $\dot \omega_{i} $ denote linear acceleration and angular acceleration and $u_{i1} ,u_{i2}\in \Re^m $ are corresponding control inputs. 
\begin{equation}\label{eqn2}
\begin{array}{l}
u_{i1} = f_1 \left(u_{ix}, u_{iy}, u_{iv}\right) \\
u_{i2} = f_2 \left(u_{ix}, u_{iy}, u_{iv}\right)
\end{array}
\end{equation}

where
\begin{equation}\label{eqn3}
\begin{array}{ll}
u_{ix}=\mathlarger{\frac{1}{\sum_{j=1}^n a_{ji}}} \mathlarger{\sum_{j=1}^n} a_{ji}\Big(x_{j}\big( \left(k-g_{j}\right)T\big) -x_{i}\big( \left(k-g_{i}\right)T\big) \Big)
\end{array}
\end{equation}
\begin{equation}\label{eqn3b}
\begin{array}{ll}
u_{iy}=\mathlarger{\frac{1}{\sum_{j=1}^n a_{ji}}} \mathlarger{\sum_{j=1}^n} a_{ji}\Big(y_{j}\big( \left(k-g_{j}\right)T\big) -y_{i}\big( \left(k-g_{i}\right)T\big) \Big)
\end{array}
\end{equation}
are feedback inputs corresponding to the local interactions with respect to the agent.
\begin{equation}\label{eqn4}
u_{iv} = v_{i}\left( \left(k-g_{i}\right)T\right) 
\end{equation}
gives feedback information about velocity of the agent. $a_{ij}=a_{ji}$ for undirected graph.


It is assumed that position and velocity are constant 
in the interval $\left[ kT, \left( k+1 \right) T\right) $. Communication and input 
time delays are represented by $g_{j}T $ and $g_{i}T $ respectively. To emulate the 
situation where topology changes due to specific positions of the agents. The control laws in equation \eqref{eqn2}
$f_1 \left(u_{ix}, u_{iy}, u_{iv}\right) $  and $f_2 \left(u_{ix}, u_{iy}, u_{iv}\right) $ obey the following algorithmic steps.
\begin{enumerate}[itemsep=0mm]
\item If $step =1 $, the agents  broadcast their positions without moving. If $1<step \leq step_{min} $, the agents broadcast messages to incorporate time-delay in the system.
\item The heading of each agent is updated by the control law $f_2 \left(u_{ix}, u_{iy}, u_{iv}\right) $ based on feedback from other agents when the agent is not inside consensus circle radius $ccr $ $(\sqrt{{u_{ix}}^2 + {u_{iy}}^2} > ccr)$ with center as average position of other agents.
\item $f_1 \left(u_{ix}, u_{iy}, u_{iv}\right) = \pm a $ if $\sqrt{{u_{ix}}^2 + {u_{iy}}^2} > ccr $ and $u_{iv} < v_{max}  $ i.e., if an agent is not inside the consensus circle radius $ccr $, set acceleration as $\pm a $ until velocity of the agent reaches $v_{max} $. 
\item $f_2 \left(u_{ix}, u_{iy}, u_{iv}\right) $ continuously updates the heading using obstacle avoidance algorithm in section-\ref{obst}. 
\item Repeat above algorithmic steps until $\sqrt{{u_{ix}}^2 + {u_{iy}}^2} \geq ccr $.
\item When an agent loses communication with other agents, the agent is brought to 
halt by setting deceleration $d $ until velocity becomes zero using $f_1 \left(u_{ix}, u_{iy}, u_{iv}\right) = -d $ if $v_i > 0 $.
\item When $\sqrt{{u_{ix}}^2 + {u_{iy}}^2} < ccr $, $f_1 \left(u_{ix}, u_{iy}, u_{iv}\right) = -d $ if $v_i > 0 $ 
and $f_1 \left(u_{ix}, u_{iy}, u_{iv}\right) = 0 $ if $v_i = 0 $ i.e., the agent is brought to halt by decelerating until velocity becomes zero. 
\item The simulation is continued till $ step_{max} $ number of steps.
\end{enumerate}

The parameters $step_{min} $, $step_{max} $, $ccr $, $a $, $d $ and $v_{max} $
are to be chosen by the users based on the scenario. The above
 position based switching algorithm without back-tracking or history following(memoryless agents) is elaborated in a flow chart as shown in \cref{flow3}.

Control inputs $u_{i1}, u_{i2}$ in (\ref{eqn2}) are able to solve 
consensus problem if the graph is a spanning tree \cite{lt08,lt13}. 
A graph becomes strongly connected when every node has
a directed path to all the other nodes. Likewise, the graph is called a spanning tree when all nodes have a 
directed path from at least one node. In distributed control, the network need not be strongly connected to reach consensus \cite{lt32}.
Consensus can be ensured if the graph is a spanning tree for fixed topology or the union of all the topologies is a spanning tree for a periodic switching topologies \cite{lt01}. However, in a spanning tree topology, non-root agents may not have information about 
other agents. This may lead to the agents not being able to decide on coming to halt 
or use back-tracking algorithm when the topology is not a spanning tree. 
To overcome the issue, we have considered strongly connected topology as a 
necessary condition.

If $x_i \rightarrow x_j, 
y_i \rightarrow y_j, \forall i \neq j$  and $\dot{x_i} \rightarrow 0, \dot{y_i}
\rightarrow 0 , \dot{\phi_i} \rightarrow 0$ , $\forall i \in \Re^n, \forall j \in \Re^n $, then consensus 
is said to be reached in position for $n$ point masses. 
It can also be said that, consensus in position of $n $ robots is reached if, $x_i \rightarrow {x_{i}}^* $ and 
$y_i \rightarrow {y_{i}}^* $ and $\dot{x_i} \rightarrow 0, \dot{y_i} \rightarrow 0 ,
\dot{\phi_i} \rightarrow 0 $ where substituting $\left({x_{i}}^*, {y_{i}}^* \right)
$ 
in the place of $\left(x_{i}, y_{i} \right) $ satisfy the condition 
$\sqrt{{u_{ix}}^2 + {u_{iy}}^2} < ccr $, $\forall i \in n $.

\section{Back-tracking Algorithm}\label{backstep}
In position based switching, incorporation of
back-tracking allows an agent to back-step when the communication 
topology is not strongly connected. The back-tracking occurs 
unless the agent reaches the position where communication is regained. At 
this point the agent takes another path and moves along the consensus point.
If an agent is not able to receive data from any of the other agents, it halts and backtracks the 
path to re-establish the communication link. Then the agent reroutes itself based on algorithmic steps, 
preserving connectedness of the network. This process is
repeated until consensus is reached.
\begin{figure}[!h]\centering
\includegraphics[height=8.5cm,width=8cm]{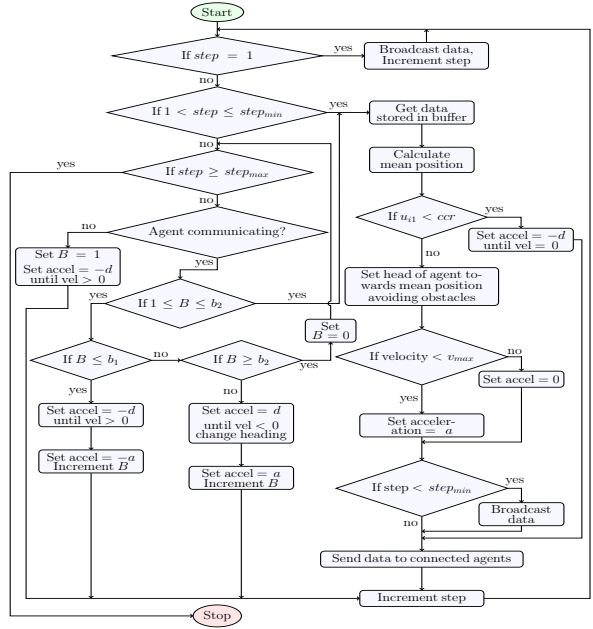}
\caption{Flow chart for position based switching with back-tracking}
\label{flow2}
\end{figure}

The algorithmic steps are as follows.
\begin{enumerate}[itemsep=0mm]
\item If $step =1 $, the agents broadcast their positions without moving. If $1<step \leq step_{min} $, the agents broadcast messages to incorporate time-delay and back-tracking in the system.
\item The heading of each agent is updated by the control law $f_2 \left(u_{ix}, u_{iy}, u_{iv}\right) $ based on feedback from other agents when the agent is not inside consensus circle radius $ccr $ $(\sqrt{{u_{ix}}^2 + {u_{iy}}^2} > ccr)$ with center as average position of other agents. 
\item $f_1 \left(u_{ix}, u_{iy}, u_{iv}\right) = \pm a $ if $\sqrt{{u_{ix}}^2 + {u_{iy}}^2} > ccr $ and $u_{iv} < v_{max}  $ i.e., if an agent is not inside the consensus circle radius $ccr $, set acceleration as $\pm a $ until velocity of the agent reaches $v_{max} $. 
\item $f_2 \left(u_{ix}, u_{iy}, u_{iv}\right) $ continuously updates the heading using obstacle avoidance algorithm in section-\ref{obst}.  
\item If an agent is not receiving data from other agents, then $f_1 \left(u_{ix}, u_{iy}, u_{iv}\right) = -d $ if $u_{iv} > 0 $ and $f_1 \left(u_{ix}, u_{iy}, u_{iv}\right) = 0 $ if $u_{iv} = 0 $
are used to bring the agent to halt. 
\item In the next step, make $f_1 \left(u_{ix}, u_{iy}, u_{iv}\right) = -a $ to back-track and it is repeated for the next $b_1$ number of steps. Then use $f_1 \left(u_{ix}, u_{iy}, u_{iv}\right) = d $ if $u_{iv} < 0 $ and $f_1 \left(u_{ix}, u_{iy}, u_{iv}\right) = 0 $ if $u_{iv} = 0 $ to bring the agent comes to halt.
\item Change the heading of agent by $\phi_b $ degrees using $f_2 \left(u_{ix}, u_{iy}, u_{iv}\right) $ and make $f_1 \left(u_{ix}, u_{iy}, u_{iv}\right) = a $ if $\sqrt{{u_{ix}}^2 + {u_{iy}}^2} > ccr $ and $u_{iv} < v_{max}  $ for next $b2 $ number of steps. 
\item Change heading towards mean position with the help of $f_2 \left(u_{ix}, u_{iy}, u_{iv}\right) $ and proceed further.
\item Iterate above steps until $\sqrt{{u_{ix}}^2 + {u_{iy}}^2} > ccr $
\item When $\sqrt{{u_{ix}}^2 + {u_{iy}}^2} \leq ccr $, the agent is brought to halt by making $f_1 \left(u_{ix}, u_{iy}, u_{iv}\right) = -d $ until $u_{iv} > 0 $ and $f_1 \left(u_{ix}, u_{iy}, u_{iv}\right) = 0 $ if $u_{iv} = 0 $.
\item The simulation is continued till $ step_{max} $ number of steps.
\end{enumerate}

Parameters $step_{min}$, $step_{max}$, $ccr$, $a$,
$v_{max}$, $b_1$, $b_2$ and $\phi_b $ are to be judicially chosen for specific
applications. The above algorithm is represented using a flow
chart in \cref{flow2}.

\section{History Following with Memory Enabled Agents}\label{memory}

\begin{figure}[!h]\centering
\includegraphics[height=8.5cm,width=8cm]{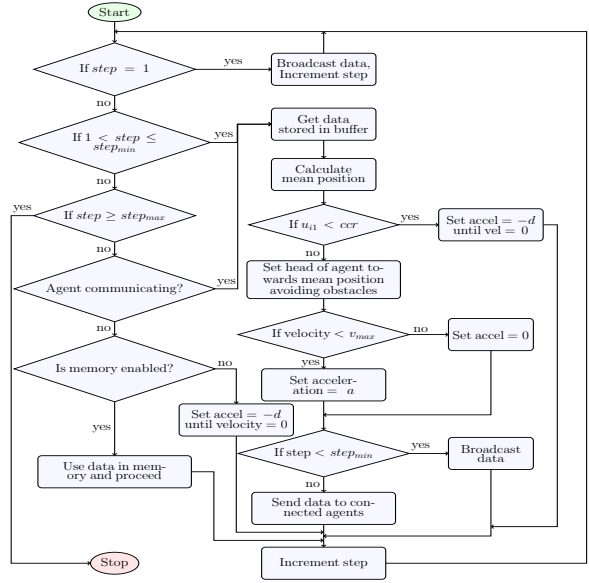}
\caption{Flow chart for position based switching with and without memory enabled agents}
\label{flow3}
\end{figure}

Here, we consider position based switching where agents can retain history of 
average consensus point. Whenever an agent loses communication with other agents, it 
proceeds with data stored in memory till it regains communication. The 
algorithmic steps are very similar to the one without back-tracking except that 
agent doesn't come to halt when it loses communication. The algorithm 
can be depicted in a flow chart as given in \cref{flow3}. Next section presents the obstacle avoidance algorithm to be incorporated with the proposed algorithms.

\section{Sensor Based Obstacle Avoidance}\label{obst}
\begin{figure}[!h]\centering
\includegraphics[height=1.5cm,width=8.5cm]{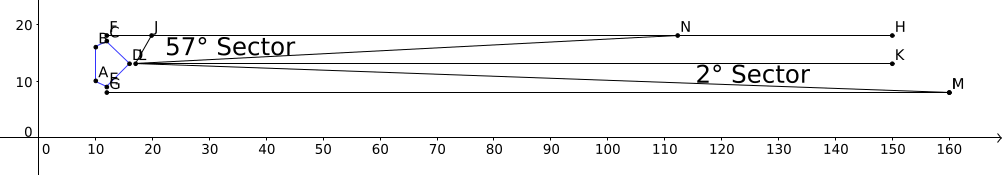}
\caption{Pictorial Representation of Obstacle Avoidance}
\label{geo}
\end{figure}

A laser sensor that is available in MRSim application is used for the purpose of 
obstacle avoidance. Beam-width of the sensor is defined to be $180^\circ $ and a 
range of $300 pixels$ is assigned. The sensor returns an array of distance values 
with which we have implemented an algorithm enumerated below. \cref{geo} gives 
a pictorial representation of the robot shape and area of obstacle avoidance which is divided 
in a number of sectors.  A detailed algorithm is as follows. 
\begin{enumerate}[itemsep=0mm]
\item Scan the
$180^\circ $ sector in the front using laser scanner to calculate cone of
avoidance. 
\item If an obstacle is found within $150 pixels $
in the sector of $\pm2^\circ $ of present heading, then turn the agent to
nearest $4^\circ $ sector with no obstacles.
\item If an
obstacle is found within $5/\sin(\theta) $ distance in the $57^\circ
$sectors of $\pm3^\circ $ to $\pm60^\circ $ of present heading, then turn
the agent to avoid obstacle with minimum of $1^\circ $ change and maximum
$30^\circ $ change.
\item If an obstacle is found within $8
pixels $ in the $30 $ sectors of $\pm61 $ to $\pm90 $, then turn the agent
by $30^\circ $.
\item If an obstacle is found within $5 pixels
$ in the sector of $\pm90^o $ of present heading, bring the agent to halt
position.
\end{enumerate}

\section{Simulation Results}\label{siml}
This section presents the simulation of multi-agent consensus for various non-switching and switching 
topologies. The simulations are done using MRSim -
Multi-Robot Simulator(V1.0) compatible on MATLAB 7.12 (R2011a)+. It can simulate upto 255 robots with user defined shape. The simulation environment is 2-dimensional 1-bit(black/white) bitmap image which
considers boarders and obstacles as zero valued pixels and rest of the
pixels as value one. The obstacles in simulation results are depicted using darker shade gray boxes and communication loss areas with lighter shade gray boxes.

\subsection{Arena and Initial Conditions}
The arena is chosen to be of size $1412 \times 773 pixels $ whose
origin is considered to be at south-west position. The arena has
boarders at $26 pixels $ in the West, at $35 pixels $ in the South, at
$1384 pixels $ in the East and at $747 pixels $ in the North. 
Each agent is fit in a circle with $5pixels $ radius. 
\begin{figure}[!h]\centering
\includegraphics[scale=0.25]{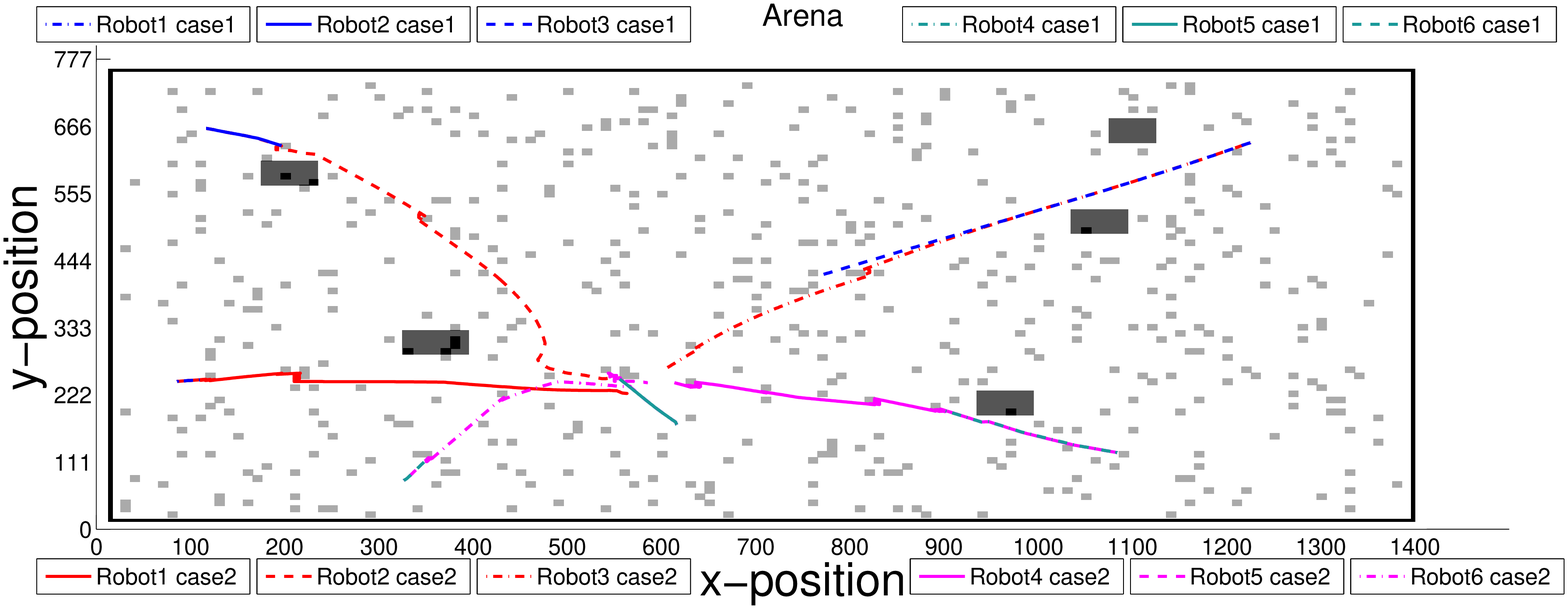}
\caption{Consensus of switching topologies based on position without back-tracking algorithm(case-$1$) and with back-tracking algorithm(case-$2$): loss of communication occurs in 5\% of the total area.}
\label{fig:4}
\end{figure}
As there are six agents, consensus circle radius $ccr $ is considered as $30 pixels $ to enable them to fit in the region even if they stand in straight line. The maximum velocity of each agent is considered as $v_{max} = 2pixels/step$, acceleration as $a = 0.1pixels/step^2 $ and deceleration as $d = 0.5pixels/step^2 $. It is considered that for the first $step_{min} =10 $ steps, all the agents broadcast data to enable back-tracking without failure. $step_{max} $ is considered as $500 $ for non-switching cases and $800 $ for position based switching cases. Number of steps for back-tracking is considered as $b_1 = 5$, number of steps for path change is considered as $b_2 = 10 $ and change in the angle after back-tracking is considered to be $\phi_b = 45^{\circ}$. 
 
\subsection{Position Based Switching Communication Topologies}
In practical cases, the position of robot decides the communication topology switching. For the sake of simplicity, 
we have assumed two communication topologies and the corresponding adjacency matrices are as given below.

$A1=\begin{bmatrix}0& 1& 0& 1& 0& 1\\ 1& 0& 1 &0& 1& 0\\ 0& 1& 0& 1& 1& 1\\ 1& 0& 1& 0& 1& 0\\ 0& 1& 0& 1& 0& 1\\1& 1& 0& 1& 1& 0 \end{bmatrix}, A2=0_{_{6 \times 6}}$

The arena will have five percent of area in which the agents lose communication and it is denoted 
with lighter shade gray boxes in the \cref{fig:4}. Simulation is done using algorithm in \cref{flow3} without memory, the agents come to halt when they lose communication, which is shown in \cref{fig:4} labelled as case-$1$. The direction robot motion is shown using a green arrow in all the simulation images. Initial positions of robots for result depicted in \cref{fig:4} are as given below,
\\
Robot1: $(86, 244)$, Robot2: $(117, 663) $, Robot3: $(1225, 639) $, Robot4: $(1084, 127) $, Robot5: $(616, 173) $, Robot6: $(326, 80) $.
\subsection{Position Based Switching Communication Topology with Back-tracking}
\begin{figure}[h]\centering
\includegraphics[scale=0.25]{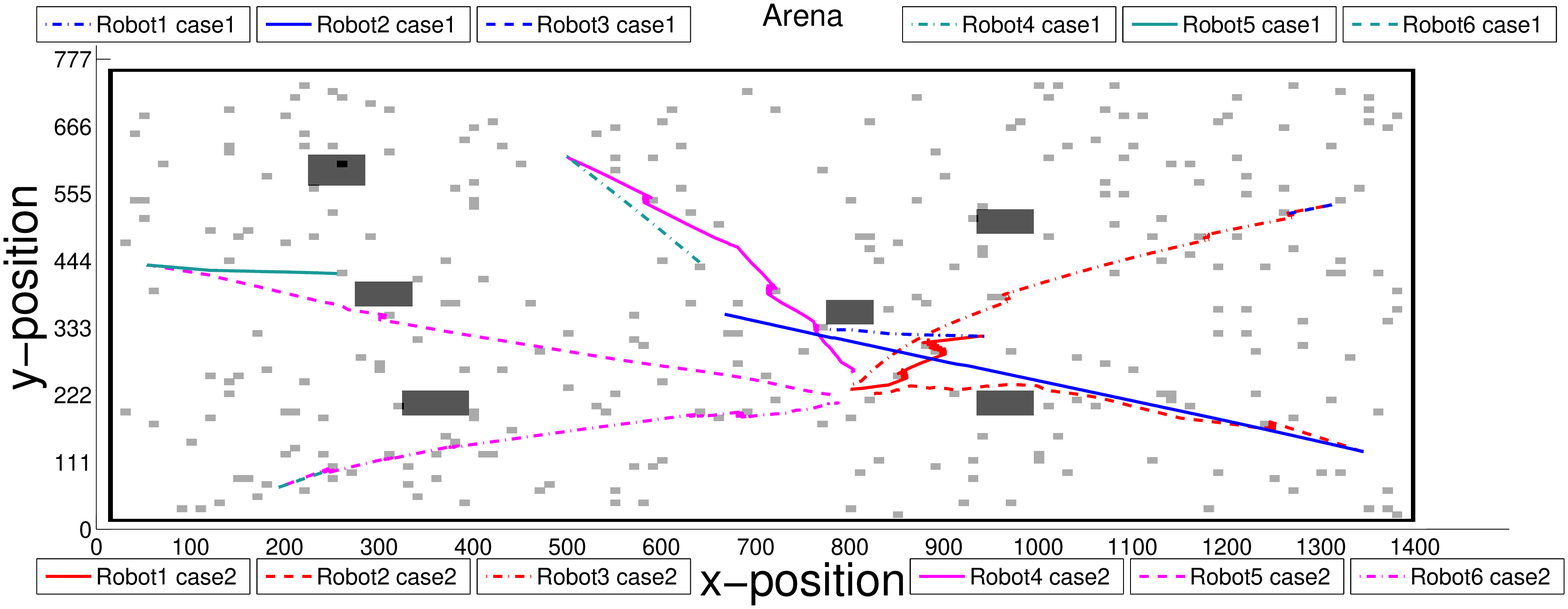}
\caption{Consensus of switching topologies based on position without back-tracking \cite{Zhou2014a}(case-1) and with back-tracking algorithm(case-2): loss of communication occurs in 3\% of the total area.}
\label{fig:8a}
\end{figure}

Here, we consider switching communication topology based on position with back-tracking by following the algorithm given in section-3. The simulations are performed using the two communication topologies whose adjacency matrices are provided earlier with time delays $g_{j}T \leq 0.6sec, g_{i}T=0.1sec$. Simulation results depict that whenever an agent is not receiving data, it reroutes using back-tracking algorithm to ensure all agents receive and send data.
Simulations are done for two different scenarios as shown in \cref{fig:4,fig:8a}. The first scenario in \cref{fig:4} has five percent of area in which the agents lose communication, the second scenario in \cref{fig:8a} has three percent of area in which the agents lose communication denoted by lighter shade gray boxes. \cref{fig:4,fig:8a} show the result where agents lose communication 
at various points but finally reach consensus using back-tracking which is labelled as case-$2$. Initial positions of robots for result depicted in \cref{fig:8a} are as given below,
\\
Robot1: $(943, 319)$, Robot2: $(1345, 128)$, Robot3: $(1312, 536)$, Robot4: $(500, 616)$, Robot5: $(54, 437)$, Robot6: $(194, 69)$

\begin{figure}[h]\centering
\includegraphics[scale=0.25]{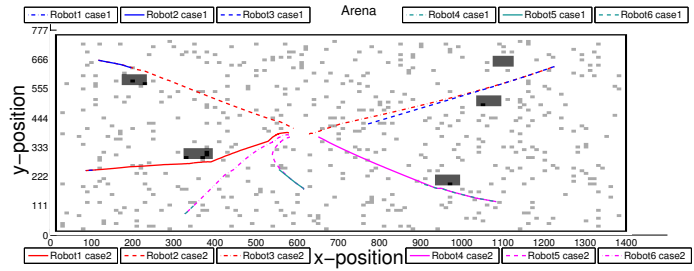}
\caption{Consensus of switching topologies based on position without memory(case-1)\cite{Zhiqiang2015} and with memory enabled agents(case-2): loss of communication occurs in 5\% of 
the total area}
\label{fig:6}
	\end{figure}
	
\subsection{Position Based Switching Network Topology with Memory Enabled Agents}
\begin{figure}[h]\centering
\includegraphics[scale=0.25]{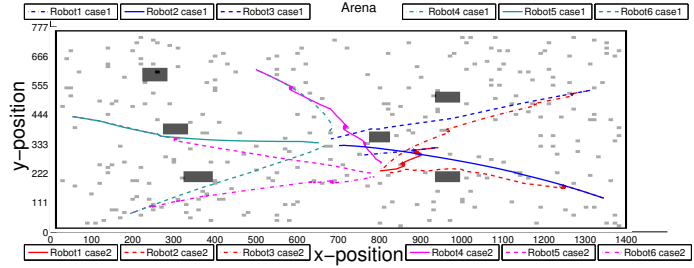}
\caption{Consensus of switching topologies based on position with memory enabled agents(case-1) and with back-tracking(case-2): loss of communication occurs in 3\% of the total area}
\label{fig:7}
\end{figure}

In this case, the simulations are performed with the two adjacency matrices as given earlier with time delays $g_{j}T \leq 0.6sec, g_{i}T=0.1sec$. Simulations are done for two different scenarios similar to earlier case as shown in \cref{fig:6} case-2 and \cref{fig:7} case-1. From \cref{fig:6}, it can be observed that the agents reach consensus but from \cref{fig:7} it can be observed that robot1 is not reaching consensus. The reason for this behaviour is, robot1 comes to halt based on data in its memory after losing communication and rest of the agents change path to avoid collision. This is a drawback which arises if an agent couldn't gain communication before rest of the agents reach consensus and come to halt. It can also be observed from \cref{fig:7} case-2, the agent reach consensus using back-tracking algorithm. Initial positions of robots for result depicted in \cref{fig:6} are as given below,
\\
Robot1: $(86, 244) $, Robot2: $(117, 663) $, Robot3: $(1225, 639) $, Robot4: $(1084, 127) $, Robot5: $(616, 173) $, Robot6: $(326, 80) $.

Initial positions of robots for result depicted in \cref{fig:7} are as given below,
\\
Robot1: $(943, 319) $, Robot2: $(1345, 128) $, Robot3: $(1312, 536) $, Robot4: $(500, 616) $, Robot5: $(54, 437) $, Robot6: $(194, 69) $.
	
\subsection{Comparison of Results}	
In the comparison shown in \cref{tb:1}, different cases have the following parameters.
{\emph{\underline{Case 1}}:} Position based switching without back tracking using algorithm in \cref{flow3} without memory.
{\emph{\underline{Case 2}}:} Position based switching without back tracking and memory using control laws in \cite{Zhou2014a} \cite{Zhiqiang2015}.
{\emph{\underline{Case 3}}:} Position based switching with back tracking using algorithm in \cref{flow2}.
{\emph{\underline{Case 4}}:} Position based switching with history following using algorithm in \cref{flow3}.
{\emph{\underline{Case A}}:}  $a=0.05$, $d=0.25$, $ V_{min}=-1$, $ V_{max}=1$, $ ccr=30$, $ step_{min}=20$, $ b_1=20$, $ b_2=20$ and arena as shown in \cref{fig:6}.
{\emph{\underline{Case B}}:}  $a=0.1$, $d=0.5$, $ V_{min}=-2$, $ V_{max}=2$, $ ccr=50$, $ step_{min}=10$, $ b_1=10$, $ b_2=20$ and arena is as shown in \cref{fig:7}.  
{\emph{\underline{Case C}}:}  $a=0.3$, $d=1$, $ V_{min}=-3$, $ V_{max}=3$, $ ccr=40$, $ step_{min}=5$, $ b_1=5$, $ b_2=15$ and arena as shown in \cref{fig:6}.
\emph{\underline{A-1, B-1, C-1}}: $3\%$ communication loss, \emph{\underline{A-2, B-2, C-2}}: $5\%$ communication loss.  

\begin{table}[!h]\centering
\caption{A comparison on the time to reach consensus for various switching topologies}
\label{tb:1}
\begin{tabular}{|m{1.85cm}|m{1cm}|m{1cm}|m{1cm}|m{1cm}|m{1cm}|m{1cm}|}
\hline
\multirow{2}{1.85cm}{\textbf{Switching topology}} & \multicolumn{6}{c|}{\textbf{Time(sec) to reach consensus}} \\
\cline{2-7}
    & \multicolumn{2}{c|}{\emph{Case A}} & \multicolumn{2}{c|}{\emph{Case B}} & \multicolumn{2}{c|}{\emph{Case C}} \\
\cline{2-7}
	& \emph{A-1} & \emph{A-2} & \emph{B-1} & \emph{B-2} & \emph{C-1} & \emph{C-2} \\   
     \hline    
\emph{Case 1} & - & - & - & - & - &\\\hline
\emph{Case 2} & - & - & - & - & - &\\\hline
\emph{Case 3} & 187.6 & 248.4 & 96.6 & 98.2 & 60.4 & 73.2 \\ \hline
\emph{Case 4} & 133 & 133 & 68.4 & 68.4 & 41.8 & 41.8 \\ \hline
\end{tabular}
\end{table}

From the simulation results shown in \cref{tb:1}, it is observed that increase in acceleration and velocity with back tracking will not directly reflect in reduced convergence time. The convergence time in a given arena is affected by initial conditions for different arenas and number of times back tracking take place for the last robot while reaching consensus. The number of times back tracked by last robot is different for all the cases, \emph{case A:} 8 and 13, \emph{case B:} 5 and 9, \emph{case C:} 6 and 7 for 3\% and 5\% loss of communication respectively. Values of $step_{min}$, $b_1$ and $b_2$ will affect the path of robot which will further affect convergence time. All the parameters should be chosen judiciously depending on size of arena, size of no communication zones and size of robots. Proposed algorithms display enough robustness to work for different set of parameters and arena conditions. As expected, observations reveal that the convergence time increases when the percentage of area with no communication increases. In the case of history following algorithm, agents are required to store the most recent update and the extra memory required is negligible. The average consensus reaching time is less in case of memory enabled agents, changes in acceleration and velocity reflect in convergence time. However, as shown in \cref{fig:7}, in some situations, following previous history may result in some error in the final destination. We should chose an algorithm considering the trade-offs between convergence time and guaranteed convergence. 

We have considered other control laws \cite{Zhou2014a}, \cite{Zhiqiang2015}, \cite{Wei2014} and the result of simulation is similar as the case without back-tracking. The authors in their control laws did not consider the complete loss of communication based on position of robots, the robots will not reach consensus once they lose communication in our arena. The resulting simulation for the control law proposed by Zhou. \emph{et al.}  \cite{Zhou2014a} is shown in \cref{fig:8a} labelled as case $1$, we can observe that the robots stop when they lose communication from other robots. Control laws with saturation is considered since the algorithms we proposed also have acceleration and velocity constraints which are similar to saturation. Most of the consensus control laws in literature are linear and switching topologies are periodic in nature. A linear control law will result in large acceleration and velocity when the agents are far from each other, those control laws cannot be used directly in our simulation since obstacle avoidance is designed for low velocity scenario and also due to limitations in hardware.  

\section{Hardware Implementation Results}\label{impl}
\begin{figure}[h]\centering
	\hspace{-0.45cm}\includegraphics[scale=0.25]{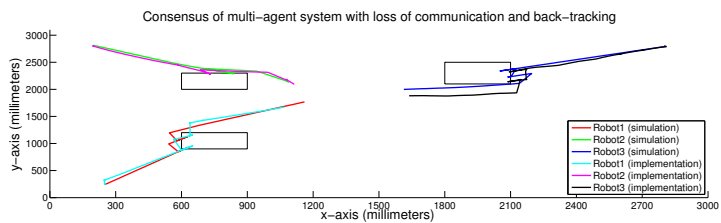}
	\caption{Position of robots in a two dimensional x-y plane}\label{fig:8}
\end{figure}

\begin{figure}[h]\centering
	\hspace{-0.45cm}\includegraphics[scale=0.25]{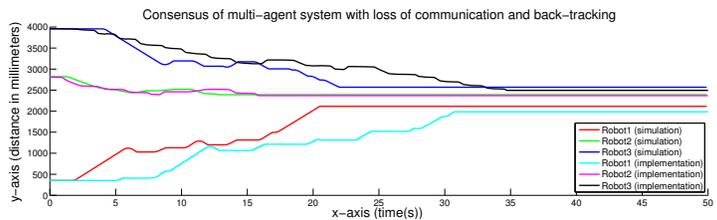}
	\caption{Distance of robots from origin with respect to time}\label{fig:9}
\end{figure}
Hardware implementation is done on a network of three agents, one is a 
Research PatrolBot manufactured by Adept MobileRobots and other two robots powered by BeagleBone-Black are developed in lab. The algorithm implemented on the system is very 
similar to simulation except the obstacle avoidance part. 

We also simulated the three agent system on MRSim for comparison of results. The 
simulation and implementation results for back-tracking case are as depicted in 
\cref{fig:8,fig:9}. The arena is chosen to be of size $3000 \times 3000 pixels $ whose origin is considered to be at south-west position where pixels in simulation are analogous to millimetres in implementation.
\cref{fig:8} shows a case with robots having 
initial positions $(250, 250), (200, 2800), (2800, 2800) $ and initial angle $0^\circ $.
For simulation, the values of acceleration and deceleration are chosen to be $50 pixels/s^2 $ and reference velocity to be $200 pixels/s $. The values of $b_1 = 5, b_2 = 10$ 
and $\phi_b = 45^{\circ} $ are chosen similar to previous simulations. For implementation, acceleration and deceleration are dynamic in nature based on PID controller output. PID controller tires to maintain a reference velocity magnitude of $200 mm/s $ in both forward and backward directions. 

\begin{figure}[h]\centering
	\hspace{-0.45cm}\includegraphics[scale=0.25]{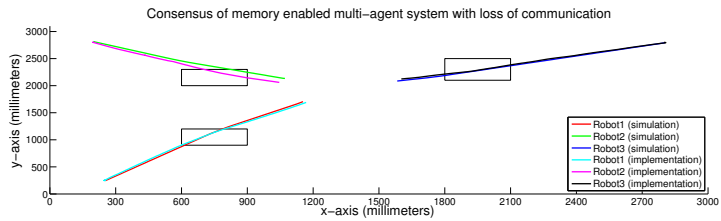}
	\caption{Position of robots in a two dimensional x-y plane}\label{fig:10}
\end{figure}

\begin{figure}[h]\centering
	\hspace{-0.45cm}\includegraphics[scale=0.25]{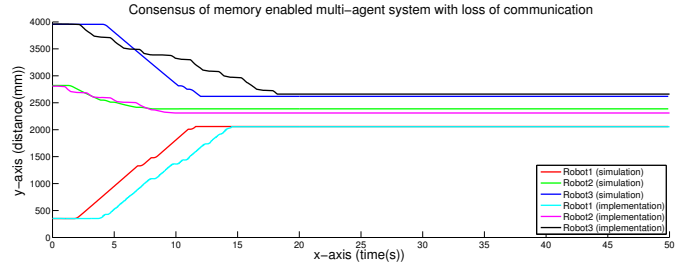}
	\caption{Distance of robots from origin with respect to time}\label{fig:11}
\end{figure}

\cref{fig:9} depicts robots' distances from origin $(0, 0)$ versus time. The rectangular boxes 
show the area where robots lose communication. The robots use 
back-tracking algorithm when they enter into that region to change its path for 
reaching destination. It is observed that there is some deviation 
in path followed and difference in convergence time between simulation and 
implementation. The path deviation is due to errors in differential drive and 
corresponding PID controller trying to maintain reference velocity of $200 mm/s $. These error is also
due to more number of robot heading corrections which increased convergence 
time.

The simulation and implementation results for memory enabled agents case are 
shown in \cref{fig:10,fig:11} with same initial conditions and arena. 
It can be observed that time taken 
to reach consensus in this scenario is less than that of back-tracking.

\section{Conclusion}\label{cncl}

Two new algorithms, to reach consensus in a multi-agent network with position based switching 
topologies, are proposed in this work. Each agent
is represented by a second order state space model where the control
input is computed using the presented algorithms. A number of static
obstacles are assumed to be present in the arena. A sensor based
obstacle avoidance algorithm ensures no collision between the agents
and the obstacles. The switching occurs depending on agents' position. 
The algorithms guarantee consensus by letting the agents to either back-track 
take take a different path or follow the previous consensus point whenever a 
loss of communication takes place. Simulation
results show that when the agents back-track and reroute whenever
they lose communication with the network, consensus is reached all the time. However, when 
they follow the previous history, situations may arise where 
some of the agents stop at locations near the actual consensus point. Otherwise, 
the time to reach consensus is less in case of history following with memory enabled 
agents. Hardware implementation without obstacles is also presented which shows the 
usefulness of the algorithms for practical purposes.

\section*{Acknowledgement}

This work is an outcome of the project ``Navigation and Path Planning
of Mobile Robots and Extension to Multi-Agent Systems'', funded by the 
Department of Science and Technology, Government of India.
\bibliographystyle{unsrt}
\bibliography{ref}

\end{document}